\documentclass[10pt, conference]{IEEEtran}
\usepackage[letterpaper, left=0.65in, right=0.65in, bottom=1.1in, top=0.75in]{geometry}
\IEEEoverridecommandlockouts
\usepackage{cite}
\usepackage{amsmath,amssymb,amsfonts}
\usepackage{graphicx}
\usepackage[caption=false,font=footnotesize]{subfig}
\usepackage{textcomp}
\usepackage{xcolor}
\usepackage{booktabs}
\usepackage{tabularx}
\usepackage{multirow}
\usepackage{diagbox}
\usepackage[binary-units]{siunitx} 
\DeclareSIUnit{\bps}{bps}
\usepackage{amsthm}
\usepackage{comment}

\def\tr{\mathrm{tr}}
\newtheorem{proposition}{Proposition}

\usepackage[ruled,vlined]{algorithm2e}
\usepackage{arydshln}

\begin{document}
\title{Achieving Optimal Performance-Cost Trade-Off in Hierarchical Cell-Free Massive MIMO}
\author{\IEEEauthorblockN{Wei Jiang\IEEEauthorrefmark{1} and Hans D. Schotten\IEEEauthorrefmark{2}}
\IEEEauthorblockA{\IEEEauthorrefmark{1}German Research Center for Artificial Intelligence (DFKI)\\Trippstadter Street 122,  Kaiserslautern, 67663 Germany\\
  }
\IEEEauthorblockA{\IEEEauthorrefmark{2}Technische Universit\"at  (RPTU) Kaiserslautern\\Building 11, Paul-Ehrlich Street, Kaiserslautern, 67663 Germany\\
 }
%\thanks{This work was supported by the German Federal Ministry of Education and Research (BMBF) through \emph{Open6G-Hub} (Grant no.  \emph{16KISK003K}) research project.}
}
\maketitle

\begin{abstract}
Cell-free (CF) massive MIMO offers uniform service via distributed access points (APs), which impose high deployment costs. A novel design called hierarchical cell-free (HCF) addresses this problem by replacing some APs with a central base station, thereby lowering the costs of fronthaul network (wireless sites and fiber cables) while preserving performance. To identify the optimal uplink configuration in HCF massive MIMO, this paper provides the first comprehensive analysis, benchmarking it against cellular and CF systems. We develop a unified analytical framework for spectral efficiency that supports arbitrary combining schemes and introduce a novel hierarchical combining approach tailored to HCF’s two-tier architecture. Through analysis and evaluation of user fairness, system capacity, fronthaul requirements, and computational complexity, this paper identifies that HCF using centralized zero-forcing combining achieves the optimal balance between performance and cost-efficiency.
\end{abstract}

\section{Introduction}

Due to its ability to offer uniform quality of service (QoS), cell-free (CF) massive multi-input multi-output (MIMO)  \cite{Ref_jiang20226GCH9} has attracted much attention in recent years. However, CF systems incur significant infrastructure costs due to the need of numerous distributed access points (APs) and its associated fronthaul network to connect these APs to a central processing unit (CPU). Recent research offers some solutions, e.g., Furtado et al. \cite{furtado2022cell} proposed cost-effective fronthaul topologies, and the authors of this paper showed that using multi-antenna APs decrease the required number of wireless sites, cutting network costs \cite{Ref_jiang2024cost}.
Wireless fronthauling offers a simple alternative to wired connections by eliminating the need for labor-intensive tasks such as trenching and fiber cable laying~\cite{demirhan2022enabling, jazi2023integrated}. Nevertheless, its adoption still presents challenges: \textit{out-of-band} implementations require dedicated transceivers and antennas, incurring hardware costs and potential licensing fees. Conversely, \textit{in-band} frequency usage sacrifices approximately half of the access link’s resources, degrading significantly performance~\cite{siddique2017downlink}. 

To address this challenge, we proposed a hierarchical cell-free (HCF) architecture as a cost-effective redesign of conventional CF systems in our previous work \cite{Ref_jiang2024hierarchical}. The core idea of this design is to aggregate a subset of distributed antennas into a central base station (cBS), which reuses the existing transceivers and antennas from decommissioned APs, thus avoiding new hardware investments.  By reducing the number of wireless AP sites and shrinking the fronthaul network scale, HCF lowers both capital and operational expenses. The results in \cite{Ref_jiang2024hierarchical} demonstrate that HCF maintains high user fairness by achieving $95\%$-likely per-user spectral efficiencies comparable to CF systems while significantly outperforming cellular networks at cell edge. However, as an initial HCF study, \cite{Ref_jiang2024hierarchical} relies on simplifications such as single-antenna APs, uncorrelated channel fading, and studies solely on centralized maximal-ratio (MR) combining without power control.

This paper presents the first comprehensive analysis of uplink HCF massive MIMO, benchmarking it against cellular and CF systems. Aimed at identifying optimal HCF configurations under practical impairments (e.g., spatially correlated fading, pilot contamination) and validating HCF’s superior balance of performance-cost trade-offs over CF and cellular systems, this work introduces three key contributions: (1) unified spectral efficiency (SE) analytical framework compatible with arbitrary combining schemes, including MR, zero-forcing (ZF), and minimum mean-squared error (MMSE); (2) a novel hierarchical combining approach tailored to HCF’s two-tier architecture; and (3) holistic analysis and evaluation to compare a series of metrics: fairness ($95\%$-likely per-user SE), system capacity, fronthaul scale, signaling overhead, and computational complexity.

The article is organized as follows. Section II offers the system models and discusses channel estimation under correlated channels and pilot contamination. Sections III and IV analyze the centralized and hierarchical combining approaches, respectively. Section V studies computational and fronthaul signaling overhead. Section VI numerical results. Finally, Section VII offers conclusions.

\section{System Model}

\begin{figure}[!t] 
    \centering
    \includegraphics[width=0.475\textwidth]{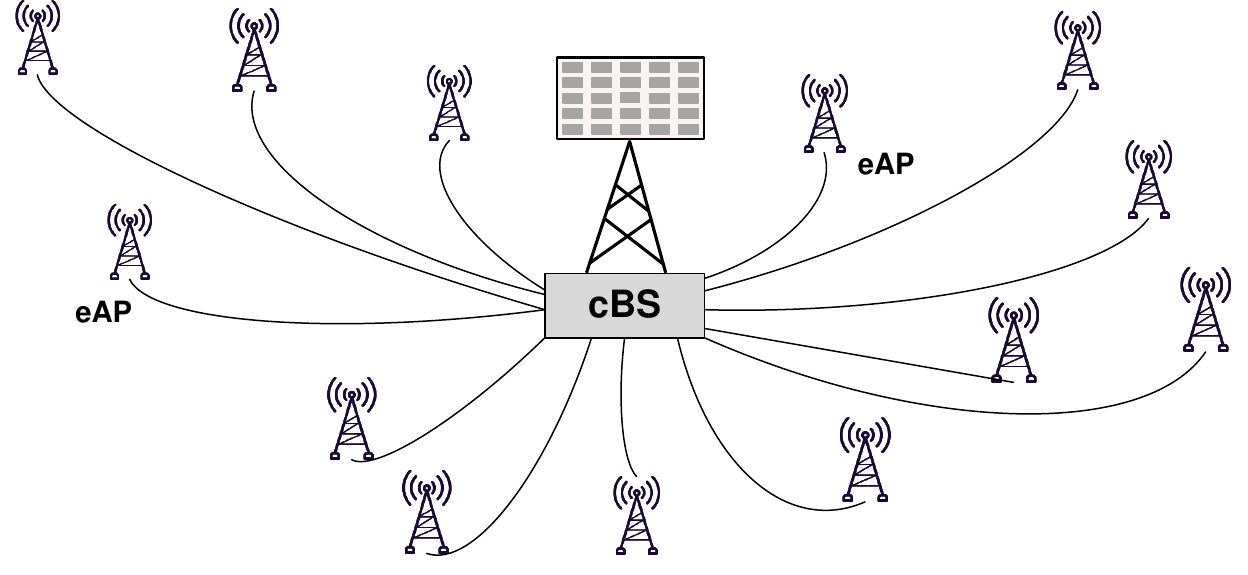} 
    \caption{An illustration of HCF: a cBS equipped with a massive antenna array, aided by distributed eAPs.}
    \label{fig:sysModel}
\end{figure}

A conventional CF system consists of $M$ service antennas to serve $K$ users. Typically, each user equipment (UE) employs a single antenna, while APs may be equipped with single or multiple antennas. A CPU coordinates all APs via a fronthaul network. To lower the scale of the fronthaul network, a hierarchical CF architecture is introduced, as shown in \figurename~\ref{fig:sysModel}, which features a cBS with $N_{b}$ antennas positioned near the center of the coverage area. The network also includes $L$ edge access points (eAPs), each equipped with $N_a$ antennas. The cBS not only serves as a central signal transceiver but also replaces the CPU in CF systems to coordinate the distributed eAPs. To maintain parity with conventional CF systems, the total number of antennas remains $N_b + L N_a = M$. The sets of indices for eAPs and users are represented by $\mathbb{L}= \{1,\ldots,L\}$ and $\mathbb{K}=\{1,\ldots,K\}$, respectively.

Assume a standard block fading model, a \textit{coherence block} spanning $\tau_c$ symbol periods, where the channel response remains approximately constant.  Using time-division duplexing, where downlink and uplink channel responses are assumed to be reciprocal, each coherence block is split into three sequential phases: $\tau_p$ channel uses for uplink training, $\tau_u$ for uplink data delivery, and $\tau_d$ for downlink data transmission such that  $\tau_c = \tau_p + \tau_u + \tau_d$.  Let $\mathbf{h}_{kl} \in \mathbb{C}^{N_a}$ represent the channel vector between eAP $l$ ($\forall l \in \mathbb{L}$) and user $k$ ($\forall k \in \mathbb{K}$). Since closely spaced antennas at a multi-antenna eAP induce correlated channel responses, channel correlation must be explicitly modeled. Consequently, the channel in each coherence block is modeled as an independent realization from a \textit{correlated} Rayleigh fading distribution: $\mathbf{h}_{kl} \sim \mathcal{CN}(\mathbf{0}, \mathbf{R}_{kl})$\cite{yu2004modeling}. Here, $\mathbf{R}_{kl}$ denotes the spatial correlation matrix that captures the channel's spatial characteristics and is defined as $\mathbf{R}_{kl} = \mathbb{E}\left[\mathbf{h}_{kl} \mathbf{h}_{kl}^H\right]$. The large-scale fading coefficient, $\beta_{kl} = \tr(\mathbf{R}_{kl}) / N_a$, encapsulates geometric path loss and shadowing effects. Similarly, let $\mathbf{h}_{k0} \in \mathbb{C}^{N_b}$ denote the channel vector between the cBS and user $k$ ($\forall k \in \mathbb{K}$), which follows $\mathcal{CN}(\mathbf{0}, \mathbf{R}_{k0})$. 

The system supports $\tau_p$ orthogonal $\tau_p$-length pilot sequences. For a scalable system, pilot sequences must be reused when the number of active users $K$ is large, i.e., $K > \tau_p$, causing \textit{pilot contamination} \cite{elijah2015comprehensive}. Let $\mathcal{P}_k$ denote the set of user indices (including UE $k$) that share the pilot sequence as UE $k$. Through linear MMSE channel estimation \cite{SIG-093}, the estimate for $\mathbf{h}_{kl}$ (and similarly $\mathbf{h}_{k0}$) is obtained as $\hat{\mathbf{h}}_{kl} =  \sqrt{p_u}  \mathbf{R}_{kl} \boldsymbol{\Gamma}_{kl}^{-1} \boldsymbol{\psi}_{kl}$, where $\boldsymbol{\Gamma}_{kl}=p_u\tau_p\sum \nolimits_{k'\in \mathcal{P}_k } \mathbf{R}_{k'l} + \sigma_z^2\mathbf{I}_{N_a}$, where $p_u$ means the power constraint of UE.

The estimation error, defined as $\tilde{\mathbf{h}}_{kl}=\mathbf{h}_{kl} - \hat{\mathbf{h}}_{kl}$, has the covariance matrix:
\begin{equation}
\boldsymbol{\Theta}_{kl}=\mathbb{E}\left[ \tilde{ \mathbf{h}}_{kl} \tilde{\mathbf{h}}_{kl}^H \right]=\mathbf{R}_{kl} - p_u \tau_p \mathbf{R}_{kl} \boldsymbol{\Gamma}_{kl}^{-1}\mathbf{R}_{kl}. 
\end{equation}
Thus, we have the following distributions:
\begin{align} \label{EQN_correlationMatrixCHest}
\hat{\mathbf{h}}_{kl} &\sim \mathcal{CN}\left(\mathbf{0}, p_u\tau_p\mathbf{R}_{kl}\boldsymbol{\Gamma}_{kl}^{-1}\mathbf{R}_{kl}\right) \\ \label{eQN_hhatdistribution}
\tilde{\mathbf{h}}_{kl} &\sim \mathcal{CN}\left(\mathbf{0}, \mathbf{R}_{kl} - p_u\tau_p\mathbf{R}_{kl}\boldsymbol{\Gamma}_{kl}^{-1}\mathbf{R}_{kl}\right).
\end{align}

For cBS channels, replace $l\rightarrow 0$, $N_a\rightarrow N_b$, and $\boldsymbol{\Gamma}_{k0} = p_u\tau_p\sum_{k'\in\mathcal{P}_k}\mathbf{R}_{k'0} + \sigma_z^2\mathbf{I}_{N_b}$. The error covariance becomes $\boldsymbol{\Theta}_{k0} = \mathbf{R}_{k0} - p_u\tau_p\mathbf{R}_{k0}\boldsymbol{\Gamma}_{k0}^{-1}\mathbf{R}_{k0}$.

\section{Centralized Uplink Combining}

In the uplink, all UEs simultaneously transmit their data symbols. Specifically, UE \( k \) transmits the symbol \( x_k \), scaled by a power coefficient \( 0 \leqslant \eta_k \leqslant 1 \). These symbols are assumed to be zero-mean, have unit variance, and are mutually uncorrelated. As a result, eAP \( l \) receives the signal
\begin{equation} \label{GS_uplink_RxsignalAP}
    \mathbf{y}_l = \sqrt{p_u} \sum\nolimits_{k\in \mathbb{K}} \sqrt{\eta_k} \mathbf{h}_{kl} x_k + \mathbf{z}_l,
\end{equation} 
where the receiver noise is modeled as \( \mathbf{z}_l \sim \mathcal{CN}(\mathbf{0}, \sigma^2_z \mathbf{I}_{N_a}) \).  
Simultaneously, the cBS receives
\begin{equation} \label{GS_uplink_RxsignalcBS}
    \mathbf{y}_0 = \sqrt{p_u} \sum\nolimits_{k\in \mathbb{K}} \sqrt{\eta_k} \mathbf{h}_{k0} x_k + \mathbf{z}_0
\end{equation} 
with \( \mathbf{z}_0 \sim \mathcal{CN}(\mathbf{0}, \sigma^2_z \mathbf{I}_{N_b}) \).
Each eAP \( l \) forwards its received pilot signals \(\boldsymbol{\Psi}_{l}\) and data signal \(\mathbf{y}_l\) to the cBS, where channel estimation and signal detection are performed.

The composite received signal at the cBS, denoted by $\mathbf{y} = [\mathbf{y}_0^T, \mathbf{y}_1^T, \ldots, \mathbf{y}_L^T]^T$, is formed by aggregating signals from $L$ eAPs. The uplink signal model can be expressed as
\begin{equation} \label{eqnUPLINKmodel}
    \mathbf{y} = \sqrt{p_u} \sum\nolimits_{k \in \mathbb{K}} \sqrt{\eta_k} \mathbf{h}_k x_k + \mathbf{z},
\end{equation}
where
\begin{itemize}
    \item $\mathbf{z} = [\mathbf{z}_0^T, \mathbf{z}_1^T, \ldots, \mathbf{z}_L^T]^T$ is the composite noise vector,
    \item $\mathbf{h}_k = [\mathbf{h}_{k0}^T, \mathbf{h}_{k1}^T, \ldots, \mathbf{h}_{kL}^T]^T$ represents the complete channel vector between user $k$ and all service antennas.
\end{itemize}
The channel vector $\mathbf{h}_k$ follows $\mathbf{h}_k \sim \mathcal{CN}\left(\mathbf{0}, \mathbf{R}_k\right)$, where  $\mathbf{R}_k$ is a block diagonal matrix $\mathbf{R}_k =\mathrm{diag}\left(\mathbf{R}_{k0}, \mathbf{R}_{k1}, \ldots, \mathbf{R}_{kL}\right)$. Through channel estimation, the cBS obtains the channel estimate $\hat{\mathbf{h}}_k$ with estimation error $\tilde{\mathbf{h}}_k = \mathbf{h}_k - \hat{\mathbf{h}}_k$, which has a covariance matrix of
$\boldsymbol{\Theta}_k = \mathrm{diag}\left(\boldsymbol{\Theta}_{k0}, \boldsymbol{\Theta}_{k1}, \ldots, \boldsymbol{\Theta}_{kL}\right)$.

To recover $x_k$, a linear detection vector $\mathbf{d}_k \in \mathbb{C}^M$ based on  $\{\hat{\mathbf{h}}_k\}_{k\in\mathbb{K}}$ is employed to compute a soft estimate:
\begin{equation}
\label{massiveMIMO:MFsoftestimateUL} 
\hat{x}_k = \mathbf{d}_k^H \mathbf{y} = \sqrt{p_u} \sum\nolimits_{k'\in\mathbb{K}} \sqrt{\eta_{k'}} \mathbf{d}_k^H \mathbf{h}_{k'} x_{k'} + \mathbf{d}_k^H \mathbf{z}.
\end{equation}
Here, $\mathbf{d}_k$ is an arbitrary detector, resulting in a unified analytical framework for performance evaluation:
\begin{proposition} \label{Proposition1}
The achievable SE for user $k$ in the uplink of HCF systems, using the centralized combining approach, is formulated as 
$R_k=(1-\frac{\tau_p}{\tau_u})\mathbb{E} [\log_2(1+\gamma_k) ]$, where the term \( (1 - \tau_p/\tau_u) \) captures the pilot overhead, and \( \gamma_k \) represents the instantaneous effective signal-to-interference-plus-noise ratio (SINR), given by  
\begin{equation} \label{GS_SINR_UL_AP}
    \gamma_k = \frac{ \eta_k | \mathbf{d}_k^H \hat{\mathbf{h}}_k |^2  }{  \mathbf{d}_k^H \left( \sum\limits_{k'\in \mathbb{K}\setminus \{k\}} \eta_{k'}  \hat{\mathbf{h}}_{k'}\hat{\mathbf{h}}_{k'}^H  + \sum\limits_{k'\in \mathbb{K}} \eta_{k'} \boldsymbol{\Theta }_{k'} +\frac{\sigma_z^2}{p_u}\mathbf{I}_M\right) \mathbf{d}_k  }.
\end{equation}
\end{proposition}
\begin{IEEEproof}
Appendix \eqref{app_a} details the derivation for \eqref{GS_SINR_UL_AP}.
\end{IEEEproof}

This paper presents three combining methods—MR, ZF, and MMSE—applied for the centralized HCF combining approach:
\begin{itemize}
    \item \textit{Centralized MR Combining:} Designed to maximize the desired signal gain, the MR combining vector for user $k$ aligns with this user’s channel estimate, i.e.,  $\mathbf{d}_{k}^{mr}=\hat{\mathbf{h}}_{k}$. While computationally efficient, it fails to suppress inter-user interference, leading to performance inferiority.
    \item \textit{Centralized ZF Combining:} Aiming to nullify inter-user interference, the ZF detector is the pseudo-inverse of the estimated channel matrix, i.e., $\mathbf{A}_{zf}=(\hat{\mathbf{H}}^H\hat{\mathbf{H}})^{-1}\hat{\mathbf{H}}^H$, where $\hat{\mathbf{H}} = [\hat{\mathbf{h}}_{1},\hat{\mathbf{h}}_{2}\ldots,\hat{\mathbf{h}}_{K}]$. Denoting $\mathbf{a}_k\in \mathbb{C}^{1\times M}$ as the $k^{th}$ row of $\mathbf{A}_{zf}$, the ZF combining vector is given by $\mathbf{d}_k^{zf}=\mathbf{a}_k^H$. 
    \item \textit{Centralized MMSE Combining:} ZF can significantly reduce interference but may amplify noise. MMSE combining strikes a balance between amplifying the desired signal and suppressing inter-user interference plus thermal noise. By minimizing the mean squared error (MSE) between \( x_k \) and \( \hat{x}_k = \mathbf{d}_k^H \mathbf{y} \), i.e., $\mathbb{E}[ |x_k - \mathbf{d}_k^H \mathbf{y}|^2 ]$, the MMSE combining vector is then obtained as
\begin{equation} \label{eq:MMSE-combining}
\mathbf{d}_{k}^{mmse}=   \left [p_u \sum\limits_{k'\in \mathbb{K}} \eta_{k'}\left( \hat{\mathbf{h}}_{k'} \hat{\mathbf{h}}_{k'}^{H} {+} \boldsymbol{\Theta }_{k'} \right) {+} \sigma_z^2  \mathbf{I}_{M} \right]^{-1}     \hat{\mathbf{h}}_{k}.
\end{equation} 
\end{itemize}

\section{Hierarchical Uplink Combining}

Beyond centralized approaches, each AP in CF systems can obtain local channel estimates and process received signals independently, using techniques like local MR \cite{Ref_ngo2017cellfree}, local ZF \cite{zhang2021local}, and local MMSE combining \cite{Ref_bjornson2020making}. Here, we design a hierarchical combining approach tailored to HCF's two-tier architecture. First, each eAP \( l \) computes estimates \(\{\hat{\mathbf{h}}_{kl}\}_{k\in \mathbb{K}}\). Then, eAP \( l \) constructs a local combining vector \( \mathbf{v}_{kl} \) for every user \( k \), and generates a soft estimate \( \hat{x}_{kl} = \mathbf{v}_{kl}^H \mathbf{y}_l \). These local estimates \( \{\hat{x}_{kl}\}_{k \in \mathbb{K}} \) are subsequently transmitted to the cBS. Meanwhile, the cBS applies its local combining vector \( \mathbf{v}_{k0} \) to its received signal \( \mathbf{y}_0 \), namely $\hat{x}_{k0} =\mathbf{v}_{k0}^H\mathbf{y}_{0}$. For detecting \( x_k \), the cBS aggregates all relevant signals using:
\begin{equation} \label{gs_uplik_softestimates_hier}
    \hat{x}_k = \hat{x}_{k0} + \sum\nolimits_{l\in \mathbb{L}} \hat{x}_{kl} = \mathbf{v}_{k0}^H\mathbf{y}_{0} + \sum\nolimits_{l\in \mathbb{L}}\mathbf{v}_{kl}^H\mathbf{y}_l.
\end{equation}
Substituting \eqref{GS_uplink_RxsignalAP} and \eqref{GS_uplink_RxsignalcBS} into \eqref{gs_uplik_softestimates_hier} yields
\begin{align} \nonumber \label{gs_ul_finalsoftestimate}
    \hat{x}_k &= \mathbf{v}_{k0}^H\left( \sqrt{p_u} \sum\nolimits_{k'\in \mathbb{K} } \sqrt{\eta_{k'}} \mathbf{h}_{k'0} x_{k'} + \mathbf{z}_0\right) \\
    &+ \sum\nolimits_{l\in \mathbb{L}}\mathbf{v}_{kl}^H\left( \sqrt{p_u} \sum\nolimits_{k'\in \mathbb{K} } \sqrt{\eta_{k'}} \mathbf{h}_{k'l} x_{k'} + \mathbf{z}_l \right).
\end{align}

In this case, the cBS is assumed to lack access to the channel estimates; otherwise, each eAP \( l \) needs to send  \(\{\hat{\mathbf{h}}_{kl}\}_{k \in \mathbb{K}}\) over the fronthaul network. Consequently, the cBS performs signal detection by means of large-scale fading decoding (LSFD), where the statistical mean \( \mathbb{E} [ \mathbf{v}_{kl}^H \hat{\mathbf{h}}_{kl} ] \) and $\mathbb{E}[\mathbf{v}_{k0}^H \hat{\mathbf{h}}_{k0}]$ serve as the approximation for the unknown channel terms.  We derive a unified analytical framework for performance evaluation, i.e.,
\begin{proposition} \label{Proposition2}
The achievable SE for user $k$ in the uplink of HCF systems, using the hierarchical combining approach, is expressed as $
    C_k=(1-\frac{\tau_p}{\tau_u}) \mathbb{E} [\log_2(1+\xi_k) ]$,
where the instantaneous effective SINR equals to
\begin{align} \label{Gs_sinrULdistributed}
    &\xi_k =\\ \nonumber
    & \frac{   \eta_k  \left | \mathbb{E}[\mathbf{v}_{k0}^H \hat{\mathbf{h}}_{k0}]+ \sum\nolimits_{l\in \mathbb{L}} \mathbb{E}[\mathbf{v}_{kl}^H \hat{\mathbf{h}}_{kl}] \right|^2   }{ \left\{   \begin{aligned}        
      & \sum\limits_{k'\in \mathbb{K}} \eta_{k'} \left(\mathbb{E}\left[\left|   \mathbf{v}_{k0}^H \hat{\mathbf{h}}_{k'0}  \right|^2 \right] + \sum\nolimits_{l\in \mathbb{L}}\mathbb{E}\left[\left|    \mathbf{v}_{kl}^H \hat{\mathbf{h}}_{k'l} \right|^2 \right] \right) \\
      &  - \eta_{k} \left(  \left|\mathbb{E}[\mathbf{v}_{k0}^H \hat{\mathbf{h}}_{k0} ] \right|^2   + \sum\nolimits_{l \in \mathbb{L}} \eta_k  \left|\mathbb{E}[\mathbf{v}_{kl}^H \hat{\mathbf{h}}_{kl} ] \right|^2  \right) \\
      & + \sum\nolimits_{k'\in \mathbb{K}} \eta_{k'}  \left(         
    \mathbf{v}_{k0}^H \boldsymbol{\Theta}_{k'0} \mathbf{v}_{k0}  + \sum\nolimits_{l\in \mathbb{L}} \mathbf{v}_{kl}^H \boldsymbol{\Theta}_{k'l}  \mathbf{v}_{kl} 
      \right) \\
      & + \frac{\sigma_z^2}{p_u} \left ( \|  \mathbf{v}_{k0} \|^2 + \sum\nolimits_{l\in \mathbb{L}} \| \mathbf{v}_{kl} \|^2 \right)
    \end{aligned} \right\}    
    }.
\end{align}
\end{proposition}
\begin{IEEEproof}
See Appendix \ref{app_c} for the proof. 
\end{IEEEproof}

Three hierarchical combining methods are applied, i.e.,
\begin{itemize}
    \item \textit{Hierarchical MR Combining:} The local MR combining vector used at eAP $l$ for user $k$ is given by $\mathbf{v}_{kl}^{mr}=     \hat{\mathbf{h}}_{kl}$, whereas at the cBS, it is $\mathbf{v}_{k0}^{mr} =    \hat{\mathbf{h}}_{k0}$.
    \item \textit{Hierarchical ZF Combining:} At eAP $l$, all local channel estimates are collectively denoted by $\hat{\mathbf{H}}_l = [\hat{\mathbf{h}}_{1l},\ldots,\hat{\mathbf{h}}_{Kl}]$. The local ZF detector is the pseudo-inverse of $\hat{\mathbf{H}}_l$, namely $\mathbf{V}_l^{zf}=\hat{\mathbf{H}}_l(\hat{\mathbf{H}}_l^H\hat{\mathbf{H}}_l)^{-1}$, such that $(\mathbf{V}_l^{zf})^H\hat{\mathbf{H}}_l^T=\mathbf{I}_K$. Denoting $\mathbf{v}_{kl}^{zf}\in \mathbb{C}^{1\times N_a}$ as the $k^{th}$ column of $\mathbf{V}_l^{zf}$, corresponding to the combining vector for user $k$ at eAP $l$. For the cBS, similarly, we have  $\mathbf{V}_0^{zf}=\hat{\mathbf{H}}_0(\hat{\mathbf{H}}_0^H\hat{\mathbf{H}}_0)^{-1}$, such that $(\mathbf{V}_0^{zf})^H\hat{\mathbf{H}}_0^T=\mathbf{I}_K$, where $\hat{\mathbf{H}}_0 = [\hat{\mathbf{h}}_{10},\ldots,\hat{\mathbf{h}}_{K0}]$. The local ZF combiner at the cBS is $\mathbf{v}_{k0}^{zf}\in \mathbb{C}^{1\times N_b}$, the $k^{th}$ column of $\mathbf{V}_0^{zf}$.
    \item \textit{Hierarchical MMSE Combining:} Similar to \eqref{eq:MMSE-combining}, we derive the local MMSE combining vector at eAP $l$ by minimizing the MSE $\mathbb{E}[ |x_k - \mathbf{v}_{kl}^H \mathbf{y}_l|^2 ]$ between \( x_k \) and its local estimate \( \hat{x}_{kl} = \mathbf{v}_{kl}^H \mathbf{y}_l \), yielding 
\begin{align} \nonumber
 \label{eq:MMSE-combining_local_eAP}
&\mathbf{v}_{kl}^{mmse} = \\  &\left[ p_u \sum\limits_{k'\in\mathbb{K}} \eta_{k'}\left( \hat{\mathbf{h}}_{k'l} \hat{\mathbf{h}}_{k'l}^{H} {+} \boldsymbol{\Theta }_{k'l} \right) {+} \sigma_z^2  \mathbf{I}_{N_a} \right]^{-1}     \hat{\mathbf{h}}_{kl}.
\end{align} 
Similarly, the combining vector at the cBS is 
\begin{align} \label{eq:MMSE-combining_local_cbs}
&\mathbf{v}_{k0}^{mmse}= \nonumber \\ 
&\left[ p_u \sum\limits_{k'\in\mathbb{K}} \eta_{k'}\left( \hat{\mathbf{h}}_{k'0} \hat{\mathbf{h}}_{k'0}^{H} {+} \boldsymbol{\Theta }_{k'0} \right) {+} \sigma_z^2  \mathbf{I}_{N_b} \right]^{-1}     \hat{\mathbf{h}}_{k0}.
\end{align} 
\end{itemize}

\section{Computational and Signaling Overhead}
To evaluate the computational complexity of various HCF combining methods, we calculate the number of complex multiplications required. For centralized MR combining, the combining vector is the channel estimate itself, incurring no additional cost. Detection \( \hat{x}_k = \hat{\mathbf{h}}_k^H \mathbf{y} \) requires \( M \) multiplications per user. For \( K \) users, the total complexity is $KM$. For centralized ZF combining, the computation of $\mathbf{A}_{zf}=(\hat{\mathbf{H}}^H \hat{\mathbf{H}})^{-1} \hat{\mathbf{H}}^H$ involves three steps: 
    \begin{itemize} 
        \item Compute \( \hat{\mathbf{H}}^H \hat{\mathbf{H}} \), requiring \( K^2 M \) multiplications.
        \item Invert \( (\hat{\mathbf{H}}^H \hat{\mathbf{H}}) \), a \( K \times K \) matrix, costing approximately \( K^3 \) multiplications (e.g., via Lower-Upper decomposition).
        \item Multiply the \( K \times K \) inverse by the \( K \times M \) matrix \( \hat{\mathbf{H}}^H \), requiring \( K^2 M \) multiplications.
    \end{itemize}
Thus, the complexity of producing $\mathbf{A}_{zf}$ is $2 K^2 M + K^3$ per coherence block. Detection \( \hat{\mathbf{x}} = \mathbf{A}_{zf} \mathbf{y} \) costs \( K M \) multiplications per channel use. Amortizing the computation of $\mathbf{A}_{zf}$ over \( \tau_u \) channel uses, the total complexity for ZF is obtained as 
    \(
    (2 K^2 M + K^3)/\tau_u + K M
    \).
Other combining methods follow analogous derivations, as provided in Tab.~\ref{tab:complexity_fronthaul}.

\begin{table}[h]
\renewcommand{\arraystretch}{1.3}
    \caption{Comparison of Computational Complexity}
    \label{tab:complexity_fronthaul}
    \centering
    \begin{tabular}{c|c|c}
        \hline \hline 
        \textbf{Method Type} & \textbf{Combining} & \textbf{Computational Complexity} \\
        \hline \hline
        \multirow{3}{*}{\shortstack[c]{\\Centralized\\HCF/CF\\ \& Cellular}}  & {MR} & $ KM $ \\
        \cline{2-3}
        & {ZF} & $ \frac{2K^2 M + K^3}{\tau_u} + KM $     \\
        \cline{2-3}
        & {MMSE} & $ \frac{M^3 + 2 K M^2}{\tau_u} + KM $   \\
        \hline
        \multirow{3}{*}{\shortstack[c]{\\Distributed\\CF}} & {local-MR} & $ KM $  \\
        \cline{2-3}
        & {local-ZF} & $ \frac{2 K^2 M + K^3M/N_a}{\tau_u} + K M $    \\
        \cline{2-3}
        & {local-MMSE} & $ \frac{MN_a^2 + 2 K MN_a }{\tau_u} + KM $   \\
        \hline
        \multirow{3}{*}{\shortstack[c]{\\HCF}} & hier-MR & $ KM $  \\
        \cline{2-3}
        & {hier-ZF} & $ \frac{2 K^2 M + K^3 \left( L + 1 \right)}{\tau_u} + K M $    \\
        \cline{2-3}
        & {hier-MMSE} & $ \frac{L (N_a^3 + 2 K N_a^2) + (N_b^3 + 2 K N_b^2)}{\tau_u} {+} K M $   \\
        \hline \hline
    \end{tabular}
\end{table}

Fronthaul overhead denotes the complex scalars exchanged between eAPs and the cBS. For centralized channel estimation, each eAP $l$ transmits $\boldsymbol{\Psi}_l$ ($N_a\tau_p$ complex scalars) and $\mathbf{y}_l$ ($N_a$ scalars/channel use), yielding $L N_a(\tau_p + \tau_u)$ per coherent block. In hierarchical approaches, eAPs instead send soft user estimates, incurring overhead of $K L \tau_u$. Comparatively, CF systems incur $M(\tau_p + \tau_u)$ (centralized) or $M\tau_u$ (local). Hence, HCF lowers overhead by a factor of $\frac{N_b}{M}$.

\section{Performance Evaluation}
Extensive numerical evaluation of the proposed HCF architecture is conducted, benchmarking its performance against CF and cellular systems with respect to a series of metrics. 
\subsection{Simulation Configuration}

The system deploys a total of $M=384$ antennas to serve $K=16$ active users. To implement HCF, by default, one-quarter of the antennas ($N_b=96$) are allocated to the cBS, along with $72$ eAPs, each equipped with four antennas, are uniformly distributed throughout the coverage area. This configuration reduces the number of wireless AP sites and  associated fiber cables by $\mathbf{25\%}$, resulting in equivalent reductions in distributed infrastructure costs of approximately $\mathbf{25\%}$. 
For a fair comparison, all network configurations maintain the same number of antennas. Thus, the CF system consists of $96$ distributed APs, each equipped with four antennas, whereas in the cellular case, a base station (BS) with $384$ antennas is used. At each simulation epoch, the locations of APs/eAPs and users are randomly varied.

The UE transmit power is limited to $p_u=200\mathrm{mW}$. At the network side, each antenna has a power constraint of $50\,\mathrm{mW}$, meaning that the BS transmits up to $19.2\,\mathrm{W}$, while each AP/eAP is restricted to $200\,\mathrm{mW}$. The system operates with a noise power spectral density of $-174\,\mathrm{dBm/Hz}$, a $9\,\mathrm{dB}$ noise figure, and a bandwidth of $5\,\mathrm{MHz}$. 
All antenna arrays are configured as half-wavelength-spaced uniform linear arrays. We simulate spatial correlation using the Gaussian local scattering model \cite[Sec.~2.6]{SIG-093}, applying an angular standard deviation of $30^\circ$. Each coherence block contains $\tau_c = 200$ channel uses (e.g., achieved by $2\,\mathrm{ms}$ coherence time and $100\,\mathrm{kHz}$ coherence bandwidth). To produce pilot contamination, we set the length of the pilot sequence at half the total number of active users, i.e., $\tau_p = 8$, where every two users share a pilot sequence.

A circular area with a macro-cell radius ($2,000$ meters) is used, where the large-scale fading is computed by $\beta=10^\frac{\mathcal{L}+\mathcal{X}}{10}$, the shadowing $\mathcal{X}\sim \mathcal{N}(0,8^2)$, and path loss uses the COST-Hata model   \cite{Ref_ngo2017cellfree}:
\begin{equation} 
    \mathcal{L}= \begin{cases}
-L_0-35\lg(d), &  d>d_1 \\
-L_0-10\lg(d_1^{1.5}d^2), &  d_0<d\leq d_1 \\
-L_0-10\lg(d_1^{1.5}d_0^2), &  d\leq d_0
\end{cases},
\end{equation}
where the three-slope breakpoints  take values $d_0=10\mathrm{m}$ and $d_1=50\mathrm{m}$ while $L_0=140.72\mathrm{dB}$, given the parameters:  the carrier frequency $1.9\mathrm{GHz}$, the height of eAP antenna $15\mathrm{m}$, and the height of UE $1.65\mathrm{m}$.

\subsection{Simulation Results}

\begin{figure*}[!tbph]
\centering
\subfloat[]{
    \includegraphics[width=0.37\textwidth]{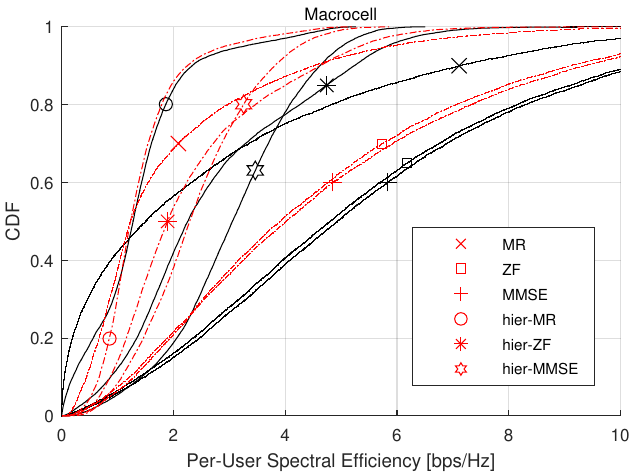}
    \label{fig:result3} 
}
\hspace{1mm}
\subfloat[]{
    \includegraphics[width=0.37\textwidth]{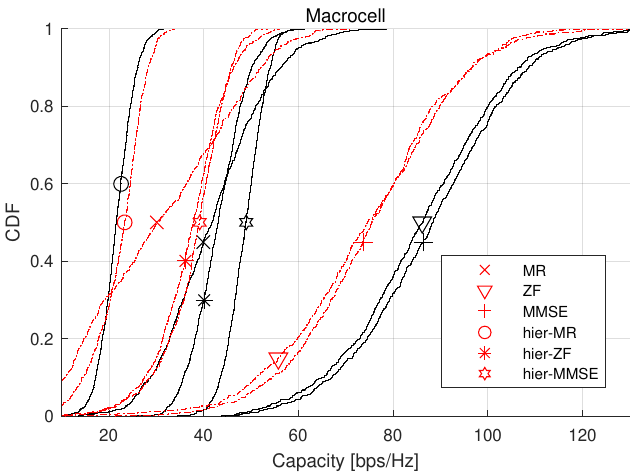}
    \label{fig:result4} 
}
\caption{Performance comparison of various HCF configurations: (a) the CDF of per-user SE; and (b) the CDF of capacity. Solid lines indicate full power transmission, while dashed-dot lines (with matching markers) represent max-min power control.}
\label{fig:resultxxx2}
\end{figure*}

\begin{figure*}[!tbph]
\centering
\subfloat[]{
    \includegraphics[width=0.37\textwidth]{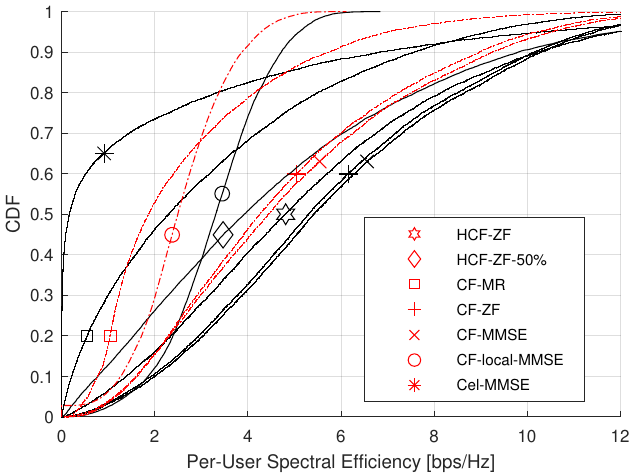}
    \label{fig:resultcompareSE} 
}
\hspace{1mm}
\subfloat[]{
    \includegraphics[width=0.37\textwidth]{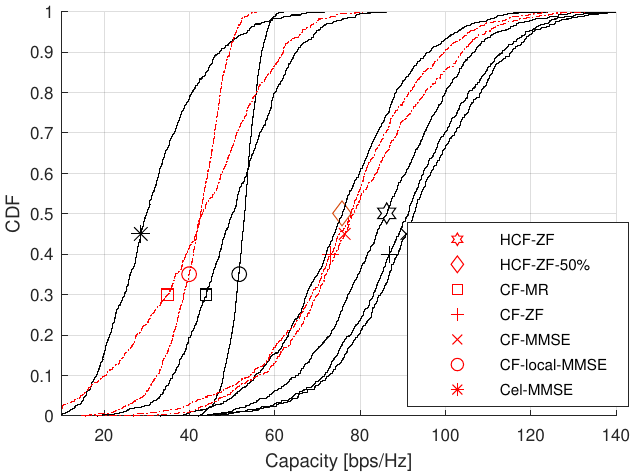}
    \label{fig:resultcompareCapacity} 
}
\caption{Performance comparison of HCF, CF, and cellular systems under macrocell scenarios: (a) the CDF of per-user SE; and (b) the CDF of capacity.}
\label{fig:resultCompare}
\end{figure*}

The simulation results, as illustrated in \figurename~\ref{fig:result3} and \figurename~\ref{fig:result4},  provide a comprehensive comparison of per-user SE and system capacity across various configurations of uplink HCF systems. The figures present the cumulative distribution functions (CDFs) under centralized (MR, ZF, MMSE) and hierarchical (hier-MR, hier-ZF, hier-MMSE) combining schemes. Configurations are distinguished by unique markers, with solid lines representing full power transmission by all users and dash-dotted lines with identical markers indicating the application of max-min power control.  We highlight two major observations:
\begin{itemize}
    \item For user fairness, ZF with full-power transmission achieves a $95\%$-likely SE of $0.81\;\mathrm{bps/Hz}$, approaching the peak value of $0.88\;\mathrm{bps/Hz}$ recorded by MMSE. Regarding to capacity, ZF and MMSE show absolute advantages over other configurations.
    \item Under max-min power control, ZF shows slightly enhanced fairness ($0.86\;\mathrm{bps/Hz}$), while MMSE's result drops to $0.78\;\mathrm{bps/Hz}$; however, system capacity degrades substantially under power control. Considering heavy computational demand of max-min optimization algorithms, this suggests that power control is not needed for ZF and MMSE.
\end{itemize}

\begin{table}[h] 
    \caption{Numbers of Complex Multiplications}
    \label{tab:complexity_summary}
    \centering
    \begin{tabular}{|c|c|c|c|}
        \hline \hline 
        \textbf{Method Type} & \textbf{MR} & \textbf{ZF} & \textbf{MMSE} \\
        \hline
        Centralized HCF/CF \& Cellular & 6,144 & 7,189 & 325,632 \\
        \hline
        Distributed CF & 6,144 & 9,216 & 6,432 \\
        \hline
        HCF & 6,144 & 8,726 & 12,504 \\
        \hline \hline
    \end{tabular}
\end{table}
The computational complexity results are detailed in Tab. \ref{tab:complexity_summary}.  Centralized ZF emerges as a computationally efficient choice for resource-limited systems, requiring 45× fewer operations than centralized MMSE and outperforming distributed and hierarchical ZF. While MR has lower complexity, centralized ZF is only $17\%$ more demanding. Regarding signaling overhead, HCF is significantly less than CF systems. Within HCF, centralized combining is far more efficient than hierarchical combining, using just $26\%$ of the overhead (e.g., 57,600 vs. 221,184 complex scalars).

\figurename \ref{fig:resultCompare} compares the per-user SE and system capacity of HCF systems against benchmarks. To simplify the visualization, HCF-ZF (centralized ZF \textbf{without} power control) is chosen to represent HCF, as it offers the best balance between performance and complexity. The benchmarks include centralized MR, ZF, and MMSE methods in CF systems, evaluated under both full power transmission (solid lines) and max-min power optimization (dash-dotted lines). Additionally, for the cellular and local CF systems, only MMSE—the best-performing configuration—is considered as a representative. To observe the impact of antenna aggregation, we use an alternative configuration, denoted by HCF-ZF-50\%, where a half of antennas are centralized at the cBS. In other words, $192$ co-located antennas are placed at the cBS, with $48$ four-antenna eAPs. This configuration further reduces the fronthaul scale to $\mathbf{50\%}$ of that in the original CF network, thereby \textbf{halving} the deployment costs. In terms of fairness, HCF-ZF achieves $95\%$-likely SE of $0.80\;\mathrm{bps/Hz}$, outperforming CF-MR ($0.38\;\mathrm{bps/Hz}$) but trailing CF-ZF ($1.24\;\mathrm{bps/Hz}$) and CF-MMSE ($1.31\;\mathrm{bps/Hz}$), while HCF-ZF-50\% has a comparable fairness ($0.38\;\mathrm{bps/Hz}$) with CF-MR. For median ($50^{th}$ percentile) capacity, HCF-ZF delivers $86.15\;\mathrm{bps/Hz}$, slightly below CF-ZF ($91.27\;\mathrm{bps/Hz}$) and CF-MMSE ($92.60\;\mathrm{bps/Hz}$), while HCF-ZF-50\% receives ($75.81\;\mathrm{bps/Hz}$). %In summary, HCF-ZF retains ${\sim}94\%$ of CF-MMSE’s capacity while lowering frauthaul scale by $25\%$, and HCF-ZF-50\% achieves ${\sim}83\%$ of maximal capacity with halved fronthaul scale. 

\section{Conclusions}

This study identified, through analysis and evaluation, that \textbf{centralized ZF combining without power control} is the best configuration for uplink HCF systems. It offers several important advantages:
\begin{itemize}
    \item \textit{Simplified Infrastructure:} HCF significantly reduces the cost of the fronthaul network (including wireless sites and fiber connections) compared to CF systems. Centralized HCF combining minimizes signaling overhead, even compared to its hierarchical combining variants.  
    \item \textit{Competitive Performance:} Centralized ZF provides near-optimal performance in terms of both  fairness and capacity, closely approaching that of centralized MMSE. 
    \item \textit{Low Complexity:} Its complexity is similar to MR, substantially less than MMSE’s high $\mathcal{O}(M^3)$ computational burden.    
    \item \textit{No Power Control Needed:} Since it doesn’t require max-min power control—a process that’s computationally intensive—the system becomes faster to operate.
\end{itemize}
In summary, HCF strike an effective balance between performance and cost-efficiency, making it a sustainable and scalable choice for future network designs.

\appendices
\section{Proof of Proposition \ref{Proposition1}} \label{app_a}
To facilitate the derivation, \eqref{massiveMIMO:MFsoftestimateUL} is further expanded as 
\begin{align} \label{HCFmassiveMIMO:MFsoftestimateUL} \nonumber
    \hat{x}_k &= \underbrace{ \sqrt{p_u \eta_k} \mathbf{d}_k^H \hat{\mathbf{h}}_k  x_k }_{\mathcal{S}_0:\:desired\:signal} + \underbrace{ \sqrt{p_u \eta_k} \mathbf{d}_k^H \tilde{\mathbf{h}}_k  x_k }_{\mathcal{I}_1:\:channel\:estimation\:error} \\&+\underbrace{\sqrt{p_u}\sum_{k'\in \mathbb{K} \backslash \{k\}} \sqrt{\eta_{k'}} \mathbf{d}_k^H \mathbf{h}_{k'}  x_{k'}}_{\mathcal{I}_2:\:inter-user\:interference}+\underbrace{\mathbf{d}_k^H\mathbf{z}}_{\mathcal{I}_3:\:colored\:noise}.
\end{align}
Given the independence of the data symbols, channel estimates, and estimation errors, the interference terms \(\mathcal{I}_1\), \(\mathcal{I}_2\), and \(\mathcal{I}_3\) are mutually uncorrelated.  Therefore, the effective SINR is given by
\begin{equation} \label{cfmmimo:formularSNR}
\gamma_k = \frac{|\mathcal{S}_0|^2}{\mathbb{E}\left[|\mathcal{I}_1 + \mathcal{I}_2 + \mathcal{I}_3|^2\right]} = \frac{|\mathcal{S}_0|^2}{\mathbb{E}\left[|\mathcal{I}_1|^2\right] + \mathbb{E}\left[|\mathcal{I}_2|^2\right] + \mathbb{E}\left[|\mathcal{I}_3|^2\right]}.
\end{equation}

As $\{\hat{\mathbf{h}}_{k0}\}_{k\in \mathbb{K}}$ and $\{\hat{\mathbf{h}}_{kl}\}_{k\in \mathbb{K}, l\in \mathbb{L}}$, as well as $\{\mathbf{d}_k\}_{k\in \mathbb{K}}$, are \textit{deterministic} for the cBS, see \cite{SIG-093}, the conditional variance of $\mathcal{I}_1$ is  derived as
\begin{align}  \nonumber
    \mathbb{E}\left[|\mathcal{I}_1|^2\right] & =  p_u \eta_k\mathbb{E}[|  \mathbf{d}_k^H \tilde{\mathbf{h}}_k  |^2]=  p_u \eta_k\mathbf{d}_k^H \mathbb{E}[   \tilde{\mathbf{h}}_k \tilde{\mathbf{h}}_k^H ]\mathbf{d}_k  \\ \nonumber 
    &= p_u \eta_k\mathbf{d}_k^H \boldsymbol{\Theta}_k\mathbf{d}_k.
\end{align}
Next, we derive
\begin{align}  \nonumber
    \mathbb{E}\left[|\mathcal{I}_2|^2\right] & \overset{(a)}{=}  p_u \mathbb{E}\left[  \sum\nolimits_{k'\in \mathbb{K} \backslash \{k\}} \left| \sqrt{\eta_{k'}} \mathbf{d}_k^H \mathbf{h}_{k'} \right |^2\right] \\ \nonumber & \overset{(b)}{=}  p_u   \sum\nolimits_{k'\in \mathbb{K} \backslash \{k\}} \eta_{k'} \mathbb{E}\left[\left| \mathbf{d}_k^H \mathbf{h}_{k'} \right |^2\right] \\ \nonumber &\overset{(c)}{=} p_u\sum\nolimits_{k'\in \mathbb{K}\setminus \{k\}} \eta_{{k'}} \mathbf{d}_k^H (\hat{\mathbf{h}}_{k'} \hat{\mathbf{h}}_{k'}^H { + } \mathbb{E}[\tilde{\mathbf{h}}_{k'} \tilde{\mathbf{h}}_{k'}^H]) \mathbf{d}_k  \\ &= p_u\sum\nolimits_{k'\in \mathbb{K}\setminus \{k\}} \eta_{{k'}} \mathbf{d}_k^H (\hat{\mathbf{h}}_{k'} \hat{\mathbf{h}}_{k'}^H + \boldsymbol{\Theta}_{k'} ) \mathbf{d}_k,  
\end{align}
where the derivations $(a)$ follows from the orthonormality of symbols, i.e., $\mathbb{E}[x_{k'}^*x_{k}]=0$ for $k'\neq k$, and $\mathbb{E}[|x_{k}|^2]=1$ for any $k$; $(b)$ uses the fact that the variance of a sum of independent random variables equals the sum of their variances; and $(c)$ is based on the independence between channel estimates and their estimation errors, namely $\mathbb{E}[\hat{\mathbf{h}}_{k'} \tilde{\mathbf{h}}_{k'}]=0$ for any $k'$. The variance of $\mathcal{I}_3$ equals $  \mathbf{d}_k^H \mathbb{E}[   \mathbf{z} \mathbf{z}^H ]\mathbf{d}_k  = \sigma_z^2 \mathbf{d}_k^H \mathbf{I}_M \mathbf{d}_k$. Thus, we obtain the effective SINR as \eqref{GS_SINR_UL_AP}.

\section{Proof of Proposition \ref{Proposition2}} \label{app_c}
The SINR expression and the variance derivations follow a similar approach to those in Appendix \ref{app_a}. Due to page limit, we omit the details. 
\bibliographystyle{IEEEtran}
\bibliography{IEEEabrv,Ref_TCOM}

\end{document}